\documentclass[]{article}
\usepackage[]{emulateapj}
\usepackage[]{times}

\submitted{Submitted for publication in
Astrophysical Journal Letters 1999 Nov 8;
accepted 2000 Jan 25}
 \begin{document}
 \tolerance=10000

\lefthead{REZZOLLA, LAMB, AND SHAPIRO}
\righthead{R-MODE OSCILLATIONS IN MAGNETIC
NEUTRON STARS}

\title{$R$-MODE OSCILLATIONS IN ROTATING
       MAGNETIC NEUTRON STARS}

\author{Luciano Rezzolla\altaffilmark{1,3}, 
	Frederick K. Lamb\altaffilmark{1,2}, 
	and 
	Stuart L. Shapiro\altaffilmark{1,2,3}}
\affil{Center for Theoretical Astrophysics,
 University of Illinois at Urbana-Champaign,
	1110 W. Green Street, Urbana IL 61801}

 \begin{abstract}
We show that $r$-mode oscillations distort
the magnetic fields of neutron stars and that
their occurrence is likely to be limited by
this interaction. If the field is
$\gtrsim10^{16}\, (\Omega/\Omega_{B})\,$G,
where $\Omega$ and $\Omega_{B}$ are the
angular velocities of the star and at which
mass shedding occurs, $r$-mode oscillations
cannot occur. Much weaker fields will prevent
gravitational radiation from exciting
$r$-mode oscillations or damp them on a
relatively short timescale by extracting
energy from the modes faster than
gravitational wave emission can pump energy
into them. For example, a $10^{10}\,$G
poloidal magnetic field that threads the
star's superconducting core is likely to
prevent the $\ell=2$ mode from being excited
unless $\Omega$ exceeds $0.35\,\Omega_{B}$.
If $\Omega$ is larger than $0.35\,\Omega_{B}$
initially, the $\ell=2$ mode may be excited
but is likely to decay rapidly once $\Omega$
falls below $0.35\,\Omega_{B}$, which happens
in $\lesssim 15^{\rm d}$ if the saturation
amplitude is $\gtrsim 0.1$. The $r$-mode
oscillations may play an important role in
determining the structure of neutron star
magnetic fields.
 \end{abstract}


\altaffiltext{1}{Department of Physics,
University of Illinois at Urbana-Champaign.}
\altaffiltext{2}{Department of Astronomy,
University of Illinois at Urbana-Champaign.}
\altaffiltext{3}{National Center for
Supercomputing Applications, University of
Illinois at Urbana-Champaign.}

\section{Introduction}
\label{intro}

The properties of $r$-mode oscillations in
rotating Newtonian stars were initially
investigated by Papaloizou \& Pringle (1978),
who suggested they may explain the
short-period oscillations seen in cataclysmic
variables. Recent work demonstrating that
gravitational radiation can cause the
$r$~modes of rotating fluid stars to grow
(see, e.g., Andersson 1998; Andersson,
Kokkotas, \& Stergioulas 1998; Friedman \&
Morsink 1998; Lindblom, Owen, \& Morsink
1998) has generated renewed interest in these
modes (see Bildsten 1998; Madsen 1998; Owen
et al.\ 1998; Andersson, Kokkotas, \& Schutz
1999; Kokkotas \& Stergioulas 1999; Levin
1999; Lindblom \& Mendell 1999; Lindblom,
Mendell, \& Owen 1999; Kojima 1998; Kojima \&
Hosonuma 1999). Unlike nonaxisymmetric
density perturbations, the $r$~modes would be
unstable at an arbitrarily small angular
velocity if neutron stars were made of
perfect fluids. This increases the potential
astrophysical significance of the $r$~modes.
The studies carried out to date have
established the basic features of the
$r$-mode instability in fluid stars, but
important questions remain. These include
interaction of $r$~mode oscillations with the
stellar magnetic field and the proton flux
tubes in the core; the effects of core-crust
coupling; and the nonlinear development of
these oscillations.

Here we investigate the distortion of the
stellar magnetic field caused by $r$~mode
oscillations and the effect of this
distortion on the onset and evolution of the
$r$-mode instability when the effects of the
superconductivity of the neutron star core
are included. We show that a magnetic field
of $\sim10^{16}$~G will prevent $r$-mode
oscillations from occurring, even if the star
is spinning very rapidly. A much weaker field
will prevent $r$-mode oscillations from being
excited by gravitational wave emission or
cause them to die out as the star spins down.
A more detailed discussion of the fluid
motions produced by the $r$ modes and their
interaction with the magnetic field of the
star will be presented in a longer paper
(Rezzolla et al.\ 2000).

\section{PROPERTIES OF $R$-MODE OSCILLATIONS}
\label{pormo}

The lowest-order Newtonian velocity
perturbations produced by small-amplitude
$r$-mode oscillations of a slowly and
uniformly rotating, nonmagnetic, inviscid,
fluid star can be determined by expanding the
fluid equations in powers of the
dimensionless mode amplitude $\alpha$ and the
ratio $\Omega/\Omega_{B}$, keeping only the
lowest-order nonvanishing terms, which are of
order $\alpha$ and $(\Omega/\Omega_{B})^2$.
Here $\Omega$ is the angular velocity of the
star and $\Omega_{B} \equiv (2/3)(\pi G {\bar
\rho})^{1/2}$ is approximately the angular
velocity at which mass-shedding is expected
to occur (see Lindblom et al.\ 1998). The
Eulerian velocity perturbations that are
solutions of these linearized fluid equations
for barytropic stars are of axial type and
may be written $\delta{\vec
v}_1(r,\theta,\phi,t) = \alpha\Omega R
\left({r/R}\right)^\ell {\vec
Y}^{B}_{\ell\ell} e^{i\sigma t}$, where $R$
is the radius of the star, ${\vec
Y}^{B}_{\ell m}$ is the magnetic-type vector
spherical harmonic, and $\sigma$ is the
angular frequency of the mode in the inertial
frame; to this same order, $\delta v_1^{r}=0$
(see Provost, Berthomieu, \& Rocca 1981;
Friedman \& Morsink 1998; Lindblom et al.\
1998).

We are interested in the velocities of fluid
elements in the frame corotating with the
unperturbed fluid of the star. These
velocities can be determined by using
$\delta{\vec v}_1$ to compute the motion of a
given element of fluid. In a spherical
coordinate basis $(r,\theta,\phi)$, the
velocity of the fluid element
located at $r,\theta,\phi$ is
  \begin{equation}
 \label{tdot}
 {\dot \theta}(t,\theta,\phi) =
 \alpha \omega f(r) c_\ell
	\sin\theta (\sin\theta)^{\ell-2} 
	\cos\left(\ell\phi +  \omega t\right)\;,
	 \ \ 
 \end{equation}
 \begin{equation}
 \label{pdot}
 {\dot \phi}(t,\theta,\phi) = 
	- \alpha \omega f(r) c_\ell
	\cos\theta (\sin\theta)^{\ell-2}
	\sin \left( \ell\phi + \omega t\right)\;,
 \end{equation}
where the dot denotes the total derivative
with respect to time, $f(r)\equiv
(r/R)^{\ell-1}$ is the radial mode function,
$c_\ell \equiv (-1/2)^\ell [(\ell-1)!]^{-1}
[(2\ell+1)!/4\pi]^{1/2}$, and $\omega =
\sigma + \ell\Omega = 2\Omega/(\ell+1)$ is
the angular frequency in the corotating
frame.

We expect $r$~waves to cause fluid elements
to drift as well as gyrate, just as sound and
water waves cause fluid elements to\break

\noindent   drift as well as oscillate (see
Landau \& Lifshitz 1959, \S~64; the time
average of $\delta{\vec v}_1$ at a {\em
fixed\/} location in the corotating frame is
of course zero). We expect the drift of fluid
elements given by the velocity field
$\delta{\vec v}_1$ to be qualitatively
correct and perhaps exact to order
$\alpha^2$. (Computing the drift of fluid
elements caused by sound waves and shallow
water waves using the solutions $\delta{\vec
v}_1$ of the relevant linearized fluid
equations as we have for the $r$~waves gives
these drifts exactly to second order in the
wave amplitude; see Rezzolla et al.\ 2000.)
As Figure~1 shows, the $\phi$ and $\theta$
excursions of a fluid element during a
gyration are comparable. The drift velocity
${\vec v}_d$ produced by $\delta{\vec v}_1$
is in the azimuthal direction; its sign and
magnitude vary with $r$ and $\theta$. Fluid
elements in the northern and southern
hemispheres equidistant from the rotation
equator drift in the same direction at the
same rate, although their gyrational motions
are different.

An analytical expression for the drift
velocity of a given fluid element can be
obtained by expanding equations~(\ref{tdot})
and~(\ref{pdot}) in powers of $\alpha$,
averaging over a gyration, and retaining only
the lowest-order nonvanishing term. For the
$\ell=2$ mode, this gives ${\vec
v}_d(r,\theta,t) = \kappa_2(\theta)
\alpha^2(t) \omega(t) R (r/R)^2 {\hat\phi}$,
where $\kappa_2(\theta) \equiv (1/2)^7
(5!/\pi) (\sin^2\theta-2\cos^2\theta)$ and
${\hat\phi}$ is the unit vector in the $\phi$
direction. Hence the net displacement in the
azimuthal direction after an oscillation is
approximately $2\pi\alpha$ times the radius
of gyration. The total displacement in $\phi$
from the onset of the oscillation at $t_0$ to
time $t$ is 
 \begin{equation}
 \label{Dxphit}
 \Delta\phi(r,\theta,t) = \kappa(\theta)
 \left(\frac{r}{R} \right)^2\!\!
 \int^t_{t_0}\alpha^2(t')\omega(t') dt'\;.
 \end{equation}
Direct numerical integrations of
equations~(\ref{tdot}) and~(\ref{pdot}) show
that expression~(\ref{Dxphit}) accurately
describes the secular displacement given by
these equations, even for $\alpha = 1$.

\section{DISTORTION OF THE STELLAR MAGENTIC
FIELD} \label{gotsmf}

The $r$~modes produce fluid motions
perpendicular to any plausible stellar
magnetic field. Such motions distort the
field, increasing its energy. As we show in
\S~4, the back-reaction of the magnetic field
on the $r$~modes can be related to the
increase in magnetic field energy. At the
core temperatures $\lesssim10^{9}$~K relevant
to $r$-mode excitation in newly formed
neutron stars, the electrical conductivity of
the core is $\gtrsim 10^{27}\, {\rm s}^{-1}$
(see Lamb 1991, \S~4.2), so the diffusion
time for fields that vary on length scales
$\gtrsim 10^5\,$cm is $\gtrsim10^9$~yr, much
longer than the timescales of interest here.
The diffusion time for older neutron stars is
much longer. We therefore neglect diffusion
and compute the effect of the fluid motions
on the magnetic field in the ideal MHD limit
in which field lines are ``frozen into'' the
fluid and carried with it. The evolution of
${\vec B}$ can be computed to leading order
in $\alpha$ without considering the change in
$\rho$, because $\nabla\cdot\delta{\vec v}_1
= 0$. Here we also neglect any instability of
the magnetic field. We defer consideration of
the back-reaction of the magnetic field on
the oscillations to \S~4. With these
approximations, the evolution of the magnetic
field is given by the induction equation
$d{\vec B}/dt = \nabla \times ({\vec u}
\times {\vec B})$, where ${d}/{dt} \equiv
({\partial}/{\partial t} + {\vec\Omega}
\times {\vec r} \cdot \nabla - {\vec \Omega}
\times)$ describes the rate of change of
$\vec B$ at a fixed position in the frame
corotating with the star and ${\vec u}$ is
the fluid velocity in this frame.

The distortion of the magnetic field during
an $r$-mode gyration is given by the
induction equation with ${\vec u} =
\delta{\vec v}_1$. The perturbation of the
field during a single oscillation is ${\delta B}
\approx (dB/dt)(\pi/\Omega)$, where $(dB/dt)
\approx {\delta v}_1 B/R \approx \alpha
\Omega B$, because ${\delta v}_1$ has
magnitude $\alpha\Omega R$ and varies with
radius and latitude on the scale $R$. Hence
the distortion of the field during an
oscillation is ${\delta B} \approx \pi\alpha
B$.

%% This fig is inserted by hand because the
%% `multicol' package cannot deal with floats
%% that have the width of a single column.
 \vbox{ \vskip -0.30truecm
 \centerline{\epsfxsize=9truecm
\epsfbox{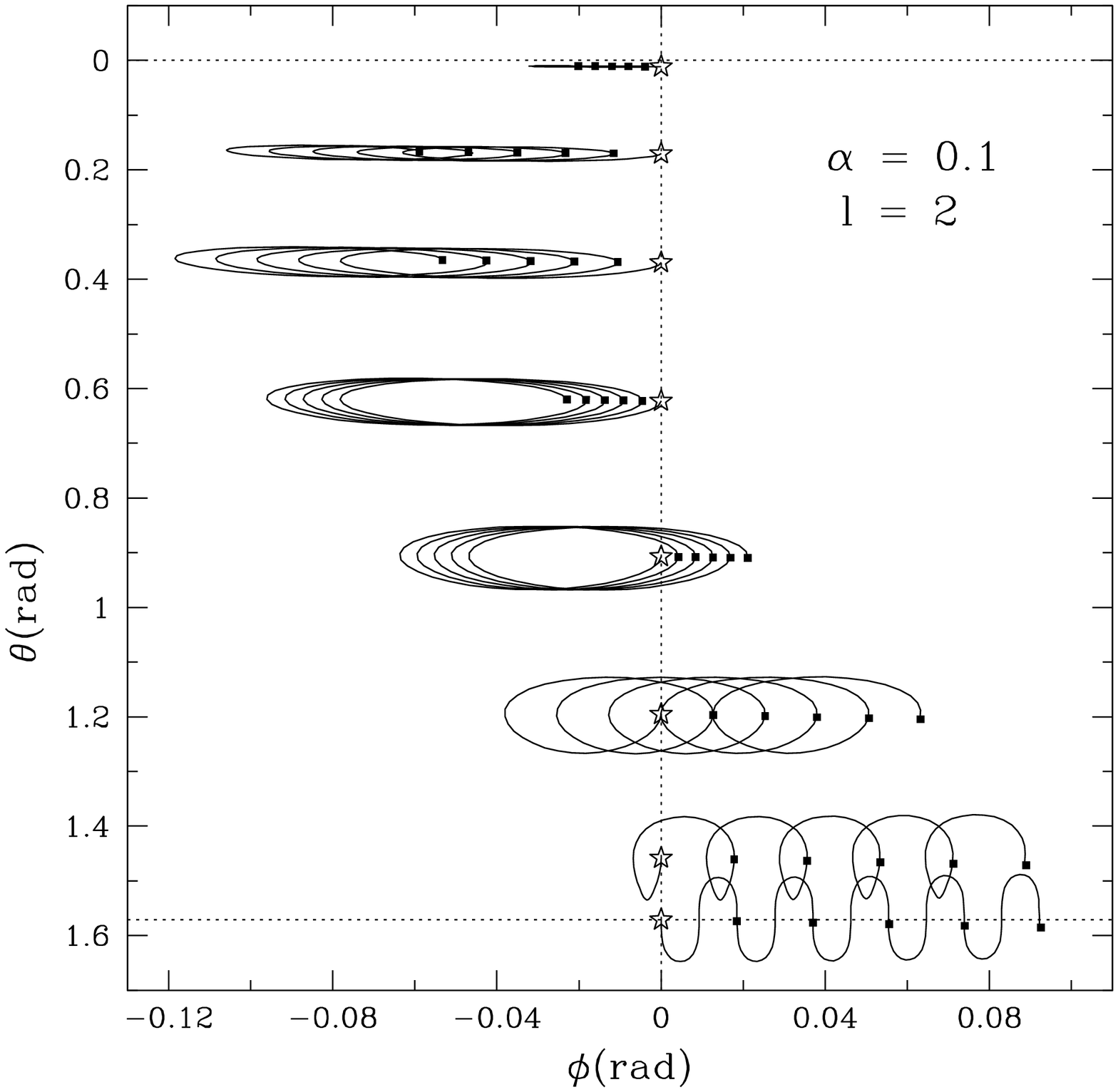}}
 \vskip -0.15truecm
 \figcaption[]{
Motions of eight fiducial fluid elements in
the northern hemisphere of a star undergoing
$\ell=m=2$ oscillations with amplitude
$\alpha=0.1$. The projected trajectories
$\theta(t) \sin\theta(t) \cos\phi(t)$ and
$\phi(t) \sin\theta(t) \cos\phi(t)$ show how
the motions over five oscillation periods
$2\pi/\omega$ would appear to a distant,
corotating observer in the rotation equator
of the star. The initial positions of the
fluid elements are indicated by the `star'
symbols; their positions after each
oscillation period are indicated by the
filled squares. The trajectories were
computed by integrating eqs.~(\ref{tdot})
and~(\ref{pdot}) numerically.
 }
 }
 \vskip 0.4truecm

The magnetic field of the star is
increasingly distorted by $r$-mode
oscillations, because ${\vec v}_d$ varies
with radius and latitude. The shear produced
by ${\vec v}_d$ generates an azimuthal
magnetic field from any poloidal field $
B_0^{\hat p}$ that is initially present. The
evolution of the azimuthal field is given by
the induction equation with ${\vec u} = {\vec
v}_d \approx \kappa \alpha^2 \omega R {\hat
\phi}$, so the change in the azimuthal field
from the time $t_0$ that the oscillation
begins to time $t$ is given approximately by
$\Delta B^{\hat\phi}(t) \approx
\int_{t_0}^{t} \kappa \alpha^2(t) \omega (t)
B_0^{\hat p} dt$. The change in $
B^{\hat\phi}$ can be computed more accurately
by noting that the induction equation can
also be written $d'{\vec B}/dt = ({\vec B}
\cdot \nabla) \delta{\vec v}$, where $d'/dt
\equiv d/dt + \delta{\vec v} \cdot \nabla$ is
the rate of change of ${\vec B}$ in the frame
comoving with the fluid. This equation can be
integrated to relate the magnetic field at
the position ${\vec x}$ of a given fluid
element at time $t$ to the magnetic field at
the position ${\vec x}_{0}$ of the same fluid
element at the earlier time $t_0$ (see Parker
1979, \S~4.3). The result is
 \begin{equation}
 \label{intgrl_indctn}
 B^j({\vec x}, t) = 
	B^k({\vec x}_{0}, t_0) 
	\frac{\partial{x^j}}{\partial x_0^k}\;,
 \end{equation}
where $\partial{x}^{j}/\partial {x}_0^k$ is
the coordinate strain between $t_0$ and $t$. 

The evolution of the azimuthal magnetic field
can be computed by averaging the azimuthal
component of equation~(\ref{intgrl_indctn})
over a time $\Delta t$ much longer than a
single gyration but much shorter than the
time for a fluid element to drift around the
star. Symmetry arguments and direct numerical
integration of equations~(\ref{tdot})
and~(\ref{pdot}) show that the
gyration-averaged displacement is independent
of $\phi$ and hence that $\partial \langle
x^\phi(t) \rangle/\partial\phi(t_0)=1$. Thus
the gyration-averaged azimuthal field at
position $\vec x$ at time $t$ is
 \begin{equation}
 \label{intgrl_indctn_phi}
 \langle
 B^{\phi}({\vec x},t)
 \rangle = 
  B^\phi_0  +  B^r_0
  \frac{
  \partial\langle{x}^{\phi}(t)\rangle}
	{\partial r_0
  	    }
 +	B^{\theta}_0
	\frac{
 \partial\langle{x}^{\phi}(t)\rangle}
	{\partial\theta_0
         }\;,
 \end{equation}
where $B^{j}_0 \equiv B^{j}({\vec x}_0,t_0)$
and the brackets indicate that the quantity
is gyration-averaged. Hereafter we will
consider only gyration-averaged quantities,
omitting the brackets for brevity.

As explained in \S~4, the damping effect of
the magnetic field can be related to the rate
at which the energy in the magnetic field
increases. After a short time, the energy in
the magnetic field is dominated by the volume
average of $|B^\phi({\vec x},t)|^2$, which
depends on the structure of the initial
magnetic field as well as on the $r$~mode in
question. $|B^\phi({\vec x},t)|$ can be
computed numerically from
equation~(\ref{intgrl_indctn}) or
analytically from
equation~(\ref{intgrl_indctn_phi}) using
equation~(\ref{Dxphit}) with $\kappa(\theta)$
replaced by its appropriately weighted
latitude average $\bar\kappa$. For the
$\ell=2$ mode and an initial magnetic field
that is dipolar and aligned with the rotation
axis of the star, $\bar\kappa=0.3$; for other
modes and field orientations, $\bar\kappa$ is
different but of the same order. Hereafter,
in computing volume averages we assume a
dipolar initial field and exclude the volume
inside $r={\scriptstyle 1\over2}R$ because of
the divergence of the dipolar field as $r
\rightarrow 0$.

We illustrate the possible effects of the
interaction of $r$~modes with the magnetic
field of a newly formed neutron star by
considering the evolutionary scenario
outlined by Owen et al.~(1998). In this
scenario the star is assumed to be completely
fluid, the initial spin rate $\Omega_0$ is
assumed to be $\Omega_{B}$, and the
saturation amplitudes of the $r$~modes are
assumed to be $\approx 1$. The velocity field
is taken to be $\delta{\vec v}_1$, even
though this expression is valid only for
$\Omega^2 \ll \Omega_B^2$ and $\alpha \ll 1$.
Owen et al.\ assumed that viscous damping is
small enough initially that gravity wave
emission can cause velocity perturbations of
$r$-mode character to grow, but that gravity
wave emission has no other effect on the
velocity field. The growth timescale is
assumed to be the gravity wave emission
timescale $|\tau_{\rm GW}|$, which is several
tens of seconds for the fastest-growing,
$\ell=2$ mode. Owen et al.\ assumed further
that exponential growth ends when the
amplitude of this mode reaches its saturation
value, that this is the only nonlinear
effect, and that the star then spins down at
the rate given by angular momentum loss to
$\ell=2$ gravitational radiation. In their
scenario, viscous damping of the $r$~modes
becomes important after about a year.

We have calculated the evolution of the
stellar magnetic field in this scenario,
except that for simplicity we have neglected
viscosity. Although we mention here our
result for $\Omega_0=\Omega_B$, in order to
allow comparison with the results of Owen et
al.\ (1998), we emphasize that the
approximations made in deriving $\delta{\vec
v}_1$ are accurate only for $\Omega \ll
\Omega_B$. A polytropic stellar model with
$\Gamma=2$ and mass $M=1.4\,M_\odot$ has
$R=12.5$~km and $\Omega_{B} = 5.60 \times
10^3$~rad~s$^{-1} \equiv \Omega_B^*$. For
$\Omega_0 = \Omega_B^*$, the $\ell=2$ mode
initially grows exponentially on the
timescale $|\tau_{\rm GW}| = 37\,$s. If the
amplitude of the initial perturbation is
$10^{-6}$ and the saturation amplitude is
0.1, the mode saturates after about 430~s.
The volume-averaged azimuthal magnetic fields
$\langle \Delta B^{\hat \phi} \rangle$ at
saturation and after 1~year are $\sim 300\,
\langle B^{\hat p}_0\rangle\,{\rm G}$ and
$\sim 10^8 \langle B^{\hat p}_0 \rangle\,{\rm
G}$, respectively, where the hats indicate an
orthonormal basis and $\langle B^{\hat p}_0
\rangle$ is the volume-averaged poloidal
component of the initial magnetic field. The
evolution of $\langle \Delta B^{\hat \phi}
\rangle$ computed analytically using
equations~(\ref{Dxphit})
and~(\ref{intgrl_indctn_phi}) agrees well
with the evolution computed numerically using
equations~(\ref{tdot}), (\ref{pdot}),
and~(\ref{intgrl_indctn}).

\section{EFFECT OF THE MAGNETIC FIELD ON
$R$-MODE OSCILLATIONS} \label{eotmf}

The evolution of $r$-mode oscillations in
magnetic stars is likely to be affected not
only by gravitational radiation and viscous
dissipation, but also by the loss of
rotational and mode energy to electromagnetic
radiation and the growth of the stellar
magnetic field. Here we focus on the
back-reaction of the magnetic field on the
$\ell = 2$ mode. Although a detailed analysis
of the change in the character and evolution
of the $r$ modes caused by their interaction
with the stellar magnetic field is beyond the
scope of the present paper, the important
qualitative effects can be estimated simply.
If the magnetic field is strong enough, it
will prevent $r$-mode oscillations from
occurring. A much weaker magnetic field will
prevent gravitational radiation from exciting
$r$-mode oscillations or cause them to die
out as the star spins down. The estimates
presented here are confirmed by more detailed
calculations (Rezzolla et al.\ 2000).

{\em Prevention of $r$-mode
oscillations}.---The distortion of the
magnetic field  during an $r$-mode
oscillation will prevent the oscillation from
occurring if the energy ${\delta E}_{m}$ that
it costs to distort the magnetic field in the
manner required by the mode is greater than
the energy ${\tilde E}$ in the mode. (The
fractional change in the mode energy produced
by gravitational wave emission during a
single oscillation period is
$4\pi/\omega|\tau_{GW}|$. It is always $\ll
1$ and therefore can be neglected in this
comparison.) We denote the critical magnetic
field that prevents $r$-mode oscillations
$B_{{\rm crit},p}$.

Consider first ${\delta E}_{m}$ for a normal
stellar core (we discuss the more realistic
but less familiar case of a superconducting
core below). An estimate of the change in the
magnetic energy of the star during an
oscillation is $\int_{V} ({\delta
B}^{2}/8\pi)\,dV$, where ${\delta B}$ is the
magnetic field perturbation. As shown in
\S~3, ${\delta B} \approx \pi\alpha B$ and
hence ${\delta E}^N_{m} \approx (\pi\alpha
B)^{2}R^{3}/30$. The mode energy ${\tilde E}$
is $\int_{V} (\rho {\delta
v}_1^{2}/8\pi)\,dV$ (see Lindblom et al.\
1998), where ${\delta v}_1 \approx \alpha R
\Omega (r/R)^4$, and hence ${\tilde E} \approx
(3/28\pi) (\alpha\Omega R)^{2}M$. Thus
${\delta E}^N_{m} > {\tilde E}$ if $B >
B^{N}_{{\rm crit},p} \approx 4 \times
10^{16}\, (\Omega/\Omega_B^*) M_{1.4}^{1/2}
R_{12.5}^{-1/2}$~G; here $M_{1.4} \equiv
M/1.4\,M_\odot$ and $R_{12.5} \equiv
R/12.5\,{\rm km}$.

The core of the neutron star is expected to
be a Type~II superconductor when the
temperature there falls below the critical
temperature $T_c \sim 3 \times 10^{9}$~K (see
Lamb 1991, \S~4.1). This is expected to occur
after a few days (see Tsuruta \& Nomoto
1988). In the superconducting core, the
magnetic field is confined to flux tubes. The
total energy of the flux tubes is $(\phi_{\rm
tot} H_{f}/8\pi)\, \ell_f$, where $\phi_{\rm
tot} \approx \pi R^2 B^{\hat p}$ is the total
magnetic flux, $H_{f}$ is the field inside a
flux tube, and $\ell_f$ is the average length
of a flux tube. During an oscillation the
flux tubes are sheared a distance $\approx
{\delta v}_1 (\pi/\omega)$, which increases
their length by ${\delta\ell} \approx (1/2)
(3\pi/2)^2 \alpha^2 (r/R)^4 R$ and their
total energy by $\delta E^{SC}_m = (1/7)
(3\pi\alpha/8)^2 B H_{f} R^3$. If $B \ll
H_{c1} \approx 10^{16}$~G, then $H_{f}
\approx H_{c1}$, but as $B \rightarrow H_{c2}
\approx 10^{17}$~G, the flux tubes crowd one
another and $H_{f} \rightarrow H_{c2}$. When
$B = H_{c2}$, the normal cores of the flux
tubes touch and the superconducting state is
destroyed (see de Gennes 1966, Chs.~3 \& 6).
If $\Omega$ is $\approx \Omega_B$, the
magnetic field needed to prevent oscillation
of the $\ell=2$ mode is $\sim 10^{16}\,{\rm
G}$ and hence $B_{{\rm crit},p} \sim
B^{N}_{{\rm crit},p}$, whether or not the
core is superconducting. If instead $\Omega
\ll \Omega_{B}$, then $B_{{\rm crit},p}^{SC}
\approx 10^{13}\, (\Omega/0.01\,\Omega_B^*)^2
H_{c1,16}^{-1}M_{1.4} R_{12.5}^{-1}\,$G,
where $H_{c1,16} \equiv H_{c1}/10^{16}\,{\rm
G}$.

{\em Damping of $r$-mode oscillations}.---The
increase in the energy of the magnetic field
produced by the $r$-mode drift reduces the
mode energy. If the rate $dE_{m}/dt$ at which
the magnetic field extracts energy from a
mode is greater than the rate $dE_{GR}/dt$ at
which gravity wave emission pumps energy into
it, the mode either will not be excited or,
if it was excited earlier, will decay. For a
given stellar magnetic field, the condition
$dE_{GR}/dt=dE_{m}/dt$ defines a critical
spin rate $\Omega_{\rm crit}$ above which
$r$~modes are excited and below which they
are damped.

If the core were normal, $dE_{m}^N/dt$ would
be $\approx (1/4\pi) \int_{V} B^{\hat\phi}
(dB^{\hat\phi}/dt)\, dV$, once $B^{\hat\phi}$
exceeds $B^{\hat p}$. As shown in \S~3,
$(dB^{\hat\phi}/dt) \approx \kappa \alpha^2
\omega (r/R)^2 B^{\hat p}$, so $dE_{m}^N/dt
\approx (\kappa/45)\alpha^{2}\Omega
(B^{\hat\phi} B^{\hat p})  R^{3}$ and the
critical spin rate would be $\Omega_{\rm
crit}^N \approx 0.3\, \Omega_B^*\,
(B^{\hat\phi} B^{\hat p}/10^{26}{\rm
G}^2)^{1/7} M_{1.4}^{-2/7} R_{12.5}^{-3/7}$,
independent of the mode amplitude. In
reality, the core is expected to be
superconducting and $dE_{m}^{SC}/dt \approx
(\phi_{\rm tot}H_{f}/8\pi) (d\ell_f/dt)
\approx (\phi_{\rm tot}H_{f}/8\pi)v_d$, once
$B^{\hat\phi}$ exceeds $B^{\hat p}$. Hence
the actual critical spin rate is likely to be
$0.7\, \Omega_B^*\, (B^{\hat p}/10^{12}{\rm
G})^{1/7} (H_f/10^{16}{\rm G})^{1/7}
M_{1.4}^{-2/7} R_{12.5}^{-3/7}$, again
independent of the mode amplitude.

\section{DISCUSSION} \label{discussion}

The results reported here indicate that
interaction of $r$-mode oscillations with the
stellar magnetic field is likely to be
important and may impose significant
constraints on the occurrence and persistence
of these oscillations in magnetic neutron
stars. Our analysis shows that if the
poloidal magnetic field of the star is
$\gtrsim 10^{10}\,$G, the increase in
magnetic energy required to keep the $\ell=2$
mode alive will exceed the energy that
gravity wave emission can supply, unless the
stellar angular velocity $\Omega$ exceeds
$0.35\, \Omega_B$. For the model of a newly
formed neutron star discussed in \S~3,
$\Omega$ falls below $0.35\,\Omega_B$ in
$\lesssim 15^{\rm d}$ if the saturation
amplitude is $\gtrsim 0.1$. Once damping of
an existing $r$~mode begins, the mode decays
quickly because gravity-wave emission
decreases rapidly as the spin rate continues
to decline.

The energy density in a magnetic field of the
strength needed to damp the $r$ modes is
$\lesssim 10^{-6}$ of the energy density in
the mode (the smallness of this number
reflects the weakness of pumping by gravity
wave emission). In contrast, the ratio of the
fluid displacement after one gyration to the
radius of gyration is $(2\pi v_d/\omega)/
(\alpha R) \approx 2\pi\kappa \alpha \approx
6\alpha$, which is many orders of magnitude
larger for mode amplitudes of interest. It
therefore seems unlikely that the magnetic
field will distort the $r$~mode velocity
field enough to cancel the fluid drift before
it suppresses the oscillation. If $r$ modes
grow to appreciable amplitudes before the
crust solidifies, their interaction with the
initial poloidal magnetic field may generate
strong azimuthal fields in young neutron
stars.

Spruit~(1998) assumed that angular momentum
loss to $r$-mode gravitational radiation
would eventually cause the interiors of old,
accreting neutron stars with millisecond spin
periods to rotate half as fast as their outer
layers. Neglecting the solid crust and its
coupling to the interior, he estimated that
this differential rotation would generate an
azimuthal magnetic field $\sim10^{17}$~G
before growth of the field ends. The results
reported here indicate that if $r$~mode
oscillations occur, then the fluid motions
they produce will themselves generate an
azimuthal magnetic field. However, the
damping effect of a moderately strong
poloidal field will prevent such oscillations
from developing.

Important issues remain unresolved. One is
whether the back-reaction of the
gravitational radiation force affects only
the amplitudes of the oscillations or also
the structure of the velocity field in the
star. We note that if it is rotating
sufficiently rapidly, a uniformly rotating,
incompressible (Maclaurin) spheroid subject
to a nonaxisymmetric perturbation is driven
by (mass quadrupole) gravitational radiation
to a nonradiating, triaxial configuration of
uniform vorticity (a ``stationary football
with internal mass currents''; see Miller
1974; Detweiler \& Lindblom 1977; Lai \&
Shapiro 1995).

Another important question is the nonlinear
development and saturation of the $r$~modes.
Do the low-$\ell$ modes grow to amplitudes of
order 1, or do they stop growing earlier
because of the loss of mode energy to
higher-$\ell$ modes? We note that pumping of
$r$~modes by gravitational radiation is a
relatively weak effect, so that even a small
energy loss rate due to mode-mode coupling
can balance it.

As the magnetic field grows, instabilities
may allow magnetic flux to leave the star,
disrupting the $r$ modes and limiting the
strength of the azimuthal field. This is most
likely in hot, young neutron stars in which
relaxation to $\beta$ equilibrium is
relatively fast and the effects of stable
stratification correspondingly reduced. Other
important issues include the effects on the
$r$ modes of the solid crust, which is
expected to begin forming when the surface
temperature falls below $\sim 3 \times
10^{10}$~K (see Lamb~1991, \S~3), and the
interaction of the neutron vortices in the
core with the proton flux tubes and the crust
(see Sauls 1989; Lamb 1991; Ruderman 1997).
Nonlinear MHD calculations and further
investigations of neutron star physics will
be needed to resolve these issues fully.

\acknowledgments     We are grateful to
D.~Markovi\'c for numerous discussions and
analyses. We thank C.~F.~Gammie, S. Hughes,
J.~C.~Miller, and S.~M.~Morsink for helpful
comments. We also thank the Aspen Center for
Physics for its warm hospitality. This
research was supported by NSF grants
AST~96-18524 and PHY~99-02833 and NASA grants
NAG~5-8424 and NAG~5-7152 at Illinois.

%\bibliographystyle{apj}
%% \bibliography

\end{document}